\documentstyle[12pt,epsfig]{article}

\newcommand{\lsim}{\mbox{$\leq$}}
\newcommand{\be}{\begin{equation}}
\newcommand{\ee}{\end{equation}}
\begin{document}  
\topmargin 0pt
\oddsidemargin=-0.4truecm
\evensidemargin=-0.4truecm
\renewcommand{\thefootnote}{\fnsymbol{footnote}}
\newpage
\setcounter{page}{0}
\begin{titlepage}   
\vspace*{-2.0cm}  
%%%\vspace*{-1.0cm}
\begin{flushright}
FT-UM-TH-03-02 \\
FIS-UM-03-10 \\
%%\vspace*{-0.2cm}
hep-ph/0304297
\end{flushright}
%\vspace*{0.5cm}
\vspace*{0.1cm}
\begin{center}
{\Large \bf
KamLAND, solar antineutrinos and the solar magnetic field} \\ 
%\vspace{0.1cm}
%\vspace{0.16cm}
\vspace{0.6cm}

\vspace{0.4cm}

{\large 
Bhag C. Chauhan$^a$\footnote{On leave from Govt. Degree College, Karsog (H P) 
India 171304. E-mail: chauhan@cfif.ist.utl.pt}, 
Jo\~{a}o Pulido$^a$\footnote{E-mail: pulido@cfif.ist.utl.pt},
 E.~Torrente-Lujan$^b$
\footnote{E-mail: e.torrente@cern.ch}}\\[2mm]
\vspace{0.15cm}
{ $^a$ {\small \sl Centro de F\'\i sica das Interac\c c\~oes Fundamentais (CFIF) \\
 Departamento de F\'\i sica, Instituto Superior T\'ecnico \\
Av. Rovisco Pais, P-1049-001 Lisboa, Portugal}\\
$^b$ {\small\sl Departamento de Fisica, Grupo de Fisica Teorica,\\ 
Universidad de Murcia, 
 Murcia, Spain}\\
}
\end{center}
\vglue 0.6truecm
\begin{abstract}

In this work the possibility of detecting solar electron antineutrinos 
produced 
by a solar core magnetic field from the KamLAND recent observations is investigated. 
We find a scaling of the antineutrino probability with respect
to the magnetic field profile in the sense that the same probability
function can be reproduced by any profile with a suitable peak field value. In 
this way the solar electron antineutrino spectrum can be unambiguosly predicted.
We use  this scaling and the negative results indicated by the 
KamLAND experiment to obtain  upper bounds 
on the solar electron antineutrino flux.   
We get 
$\phi_{\bar\nu}<3.8\times 10^{-3}\phi(^8B)$ 
%%%%or  $\phi_{\bar\nu}<5.5\times 10^{-3}\phi(^8B)$ 
at 95\% CL. 
%%assuming Gaussian or Poissonian statistics respectively. 
For 90\% CL this becomes 
$\phi_{\bar\nu}<3.4\times 10^{-3}\phi(^8B)$, 
%% and $\phi_{\bar\nu}<4.9\times 10^{-3}\phi(^8B)$,
an improvement by a factor of 3-5 with respect to existing bounds.
These limits are independent of the detailed structure of the
magnetic field in the solar interior. 
We also derive upper bounds on the peak field value which are 
uniquely determined for a fixed solar field profile. In the most
efficient antineutrino producing case, we get  (95\% CL)
an upper limit on the product of the neutrino magnetic moment by 
the solar field 
$\mu B< 2.8\times 10^{-19}$ MeV or  
$B_0 \leq 4.9 \times 10^7 G$  for $\mu_\nu=10^{-12}\mu_B$.

%In this work we find that, for a wide 
%family of magnetic field profiles in the sun 
%interior, the antineutrino appearance probability is largely determined 
%by the magnetic field intensity but not by its shape.
%From the KamLAND observations (first 145 days of data taking),  we
%conclude that an  upper limit on the product of the intrinsic 
%neutrino magnetic moment and the value of the solar core magnetic

%95\% CL (for the LMA solution) can be obtained.
%The value of the peak magnetic field at the sun core is typically
%$B_0 \leq 5.95 \times 10^7 G$ at 95\% CL for $\mu_\nu=10^{-12}\mu_B$.
%Explicit limits on neutrino transition moments are also obtained. 
%In the most conservative case, 
%$\mu\lsim 2.5\times 10^{-11}\ \mu_B$ (95\% CL).
%These limits are independent of the detailed structure of the 
%magnetic field in the solar interior.

\end{abstract}

\end{titlepage}   
\renewcommand{\thefootnote}{\arabic{footnote}}
\setcounter{footnote}{0}
%\section{Introduction}

%{\bf 1.} 
\section{Introduction}

The recent results from the KamLAND experiment \cite{Eguchi:2002dm} have 
asserted 
that the large mixing angle solution (LMA) is the dominant one for the 
34 year old solar neutrino problem SNP \cite{Akhmedov:1997yv}. 
Although neutrinos were
known, before KamLAND data, to oscillate \cite{Aliani:2002ma,klothers}, 
it was not clear if neutrino 
oscillations were the major effect underlying the 
solar neutrino deficit or whether they played any role at all.
It had been clear however that this deficit had to rely on 'non-standard'
neutrino properties. To this end, the spin flavor precession (SFP) \cite{SV,LM,Ak},
based on the interaction of the neutrino magnetic 
moment with the solar magnetic field was, second to oscillations, the most 
attractive scenario \cite{generalrandom}.    

SFP, although certainly not playing the major role in the solar neutrino deficit,
may still be present as a subdominant process, provided neutrinos have a 
sizeable transition magnetic moment. Its signature will be the appearance of
solar antineutrinos \cite{LM,Ak1,alianiantinu} 
which result from the combined effect of the vacuum
mixing angle $\theta$ and the transition magnetic moment $\mu_{\nu}$ converting 
neutrinos into antineutrinos of a different flavor. 
This can be schematically 
shown as 

\be
{\nu_e}_L \rightarrow {\nu_{\mu_L}} \rightarrow {\bar\nu_{e_R}},
\ee
\be
{\nu_e}_L \rightarrow {\bar\nu_{\mu_R}} \rightarrow {\bar\nu_{e_R}}
\ee
with oscillations acting first and SFP second in sequence (1) and in reverse
order in sequence (2). Oscillations and SFP can either take place in the same 
spatial region, or be spatially separated. Independently of their origin,
antineutrinos with energies above 1.8 MeV can be detected in KamLAND via the
observation of positrons from the inverse $\beta$-decay reaction 
$\bar\nu_{e} + p \rightarrow n+e^{+}$ and must all be originated from 
$^8 B$ neutrinos.

The purpose of this work is to relate the solar magnetic field profile to
the solar antineutrino event rate in KamLAND which is a component of 
the total positron 
event rate in the  reaction above. In a previous paper \cite{Akhmedov:2002mf}
the question of what can be learned about the strength and coordinate
dependence of the solar magnetic field in relation to the current upper 
limits on the solar $\bar\nu_{e}$ flux was addressed. The system of equations
describing neutrino evolution in the sun was solved analytically in perturbation 
theory for small $\mu_{\nu}{B}$, the product of the neutrino magnetic moment
by the solar field. The three oscillation scenarios with the best fits were 
considered, namely LMA, LOW and vacuum solutions. In particular for LMA it was 
found that the antineutrino probability depends only on the magnitude of the 
magnetic field in the neutrino production zone. Neutrinos were, in the 
approximation used, considered to be all produced at the same point 
($x=0.05R_S$), where $^8 B$ neutrino production is peaked. 
In this work we will consider the more realistic case of a convolution 
of the production distribution spectrum with the field profile in that region.
It will be seen that this convolution leads to an insensitiveness of the
antineutrino probability with respect to the solar magnetic field profile,
in the sense that different profiles can correspond to the same probability
function, provided the peak field values are conveniently scaled. As a
consequence, an upper bound on the solar antineutrino flux can be derived which
is independent of the field profile and the energy spectrum of this flux
will also be seen to be profile independent.   

Up to now, in their published data from a 145 day run, the KamLAND 
\cite{Eguchi:2002dm} experiment has observed 
a total antineutrino flux compatible with the expectations coming 
from the nearby nuclear reactors.
 Given  this fact, and evaluating the positron event rate for the above reaction 
for different solar field profiles, we will derive upper bounds for 
the peak field value in each profile. In our study we will assume the 
astrophysical upper bounds on the neutrino magnetic moment $\mu_{\nu}<
(1-3)\times 10^{-12}\mu_B$ \cite{Raffelt:pj} to be all satisfied and take 
$\mu_{\nu}= 10^{-12} \mu_B$.

Upper limits on the solar antineutrino flux, the intrinsic magnetic moment
and the magnetic field at the bottom of the convective zone were recently
obtained \cite{alianiantinu} from the published KamLAND data. Here we address
however a different antineutrino production model where the magnetic field 
at the solar core is the relevant one. 

\section{The solar antineutrino probability}

%{\bf 2.} 
We start with the probability that a ${\nu_e}_L$ 
produced inside the sun will reach the earth as a $\bar\nu_{e_R}$
\be
P(\nu_{e_L} \rightarrow \bar\nu_{e_R})=P(\nu_{e_L} \rightarrow 
\bar\nu_{\mu_R};R_S) \times 
P(\bar\nu_{\mu_R} \rightarrow \bar\nu_{e_R};R_{es})
\ee
in which the first term is the SFP probability, $R_S$ is the solar radius
and the second term is given 
by the well known formula for vacuum oscillations
\be
P(\bar\nu_{\mu_R} \rightarrow \bar\nu_{e_R};R_{es})={\sin}^{2}2\theta 
~~{\sin}^{2}\! \left(\frac{\Delta m^{2}}{4E}R_{es}\right)=\frac{1}{2}.
\ee
Here $R_{es}$ is the distance between the sun and the earth and the rest of
the notation is standard. Since $1.8 MeV < E < 15 MeV$ and, for LMA, 
$\Delta m^{2}=6.9\times10^{-5}eV^{2}$, $\sin^2 2\theta=1$ \cite {Eguchi:2002dm}, 
we take the $\bar\nu_{\mu_R} \rightarrow \bar\nu_{e_R}$ vacuum oscillations 
to be in the averaging regime.

The SFP amplitude in perturbation theory for small $\mu B$ 
is \cite{Akhmedov:2002mf} \footnote{For notation we refer the reader to 
ref. \cite{Akhmedov:2002mf}.}
\be
A(\nu_{e_L} \rightarrow \bar\nu_{\mu_R})= \frac{\mu B(r_i) \sin^{2}\theta (r_i)}{g^{'}_2(r_i)}.
\ee
A key observation is that the antineutrino appearance probability is 
dependent on the production point of its parent neutrino 
so that the overall antineutrino probability is
\be
P(\nu_{e_L} \rightarrow \bar\nu_{e_R})= \frac{1}{2} \int |A(\nu_{e_L} 
\rightarrow \bar\nu_{\mu_R})|^2 f_{B}(r_i)dr_i
\ee
where $f_{B}$ represents the neutrino production distribution function 
for Boron
neutrinos \cite{Bahc} and the integral extends over the whole production 
region. As shall be seen, owing to this integration, the energy shape of 
probability (6) is largely insensitive to the magnetic field profile.  

The positron event rate in the KamLAND experiment originated from 
solar antineutrinos is then
\be
S=Q_0\int_{E_e^0}^{\infty} dE_e \int_{E_m}^{E_M} \epsilon (E^{'}_e) R(E_e,E^{'}_e) 
\phi_{\bar\nu}(E) \sigma(E)dE.
\label{erate}
\ee
In this expression $Q_0$ is a 
normalization factor which takes into account the number of atoms of the 
detector and its live time exposure \cite{Eguchi:2002dm} and 
$E$ is the antineutrino energy, related to the physical positron energy 
by $E^{'}_e=E-(m_N-m_P)$ 
to zero order in $1/M$, the nucleon mass. We thus
have $E_m=1.804MeV$, while the KamLAND energy cut is $E_e^0=2.6MeV$. 
The functions
$\epsilon$ and $R$ denote the detector efficiency and the Gaussian 
energy resolution function of the detector
\be
R(E_e,E^{'}_e)=\frac{1}{s \sqrt{2\pi}}\exp 
\left[\frac{-(E_e-E^{'}_e)^2}{2 s^2}\right].
\ee
In our analysis we use for the energy resolution in the prompt 
positron detection the expression $s(E_e)=0.0062+0.065\surd E_e$ 
with all energy units in MeV. This is obtained from the raw calibration 
data presented in Ref.\cite{klstony}. Moreover, we assume a  
408 ton fiducial mass and  the detection efficiency is 
taken independent of the energy \cite{klstony},
 $\epsilon\simeq 80\%$, which amounts to 162 ton$.$ yr of antineutrino 
data.
The antineutrino cross section $\sigma(E)$ was taken from ref.\cite{Vogel:1999zy}
and we considered energy bins of size $E_e=0.425\ MeV$ 
in the KamLAND observation range $(2.6-8.125)~MeV$ \cite{Eguchi:2002dm}.
The antineutrino spectral flux $\phi_{\bar\nu} (E)$ in eq. \ref{erate}
can be written as  $\phi_{\bar\nu} (E)=\phi_{\bar\nu}^0\times f(E)$ 
where $\phi_{\bar\nu}^0$ is the total antineutrino flux and 
$f(E)$ is some function of the energy normalized to one.
The function $\phi_{\bar\nu} (E)$ is on the other hand a simple product of 
the Boron neutrino spectral flux 
$\phi_B(E)$ which can be found in Bahcall's homepage \cite{Bahc}
and the antineutrino appearance probability we obtained above: 
 $\phi_{\bar\nu} (E)=\phi_B (E)\times  P(E)$.
The almost insensitivity of the shape of $P(E)$ to the shape of the magnetic field
profile is thus necessarily reflected in $\phi_{\bar\nu} (E)$.
The only significant dependence appears on the normalization constant 
$\phi_{\bar\nu}^0$ which is essentially proportional to the square 
of the magnetic field at the solar core. We make use of this 
behavior to obtain, for each given profile, upper limits on the core magnetic 
field, the total antineutrino flux and the intrinsic neutrino magnetic moment.

%\subsection{Solar Magnetic Profiles}

\begin{figure}[h]
\setlength{\unitlength}{1cm}
\begin{center}
%\hspace*{-1.8cm}
\hspace*{-1.6cm}
\epsfig{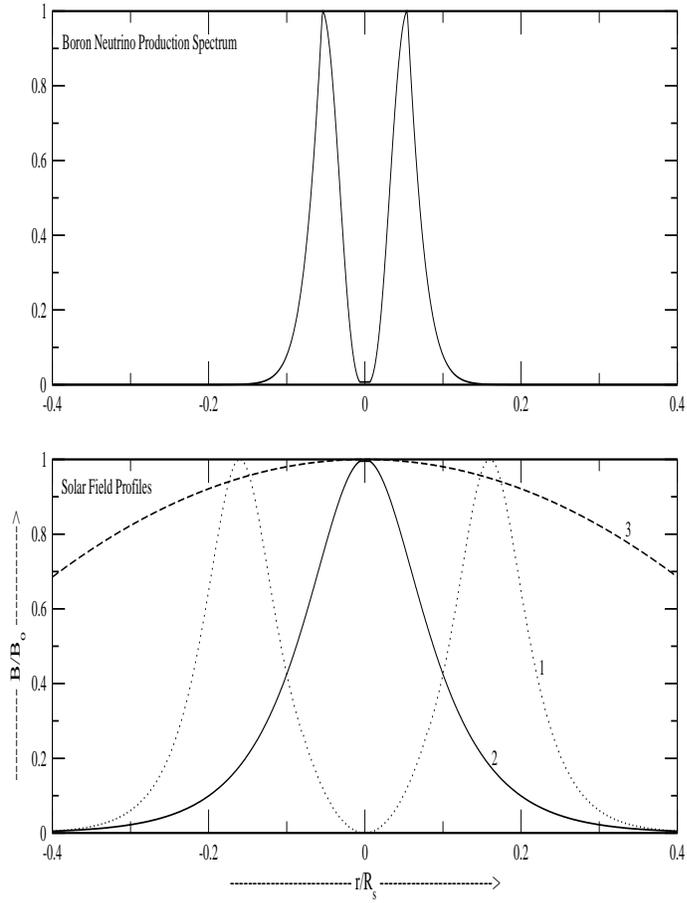}
\end{center}
\caption{ \it Upper panel: $^8B$ neutrino production spectrum (in arbitrary 
units) as a function of the radial coordinate. Lower panel: the three solar field 
profiles considered in the main text normalized to $B_0$, the peak field value.}  
\label{fig1}
\end{figure}

%{\bf 3.}
As mentioned above, for the LMA solution only the solar field profile in the
neutrino production region \cite{Akhmedov:2002mf} can affect the antineutrino 
flux. Hence
we will discuss three profiles which span a whole spectrum of possibilities at 
this region.
We study from a vanishing field (profile 1) to a maximum field at the solar center, with, 
in this second case, either a fast decreasing field intensity (profile 2) or a 
nearly flat one (profile 3) in the solar core 
(see fig. \ref{fig1}, lower panel). 
Thus, we consider 
respectively the following three profiles

{\it Profile 1}
\be 
B(r)=B_0[\cosh(9r)-1]~~,~~|r| \leq r_c
\ee 
\be
B(r)=B_0/\cosh[25(r-r_R)]~~,~~|r|>r_c,
\ee
with $r_c=0.08$, $r_R=0.16$,

{\it Profile 2}
\be
B(r)=B_0/\cosh(15r)~~,~~|r| \geq 0,
\ee

{\it Profile 3}
\be
B(r)=B_0[1-(r/r_c)^2]~~,~~|r| \leq r_c,
\ee
with $r_c=0.713$.

We
also show in fig. \ref{fig1} (upper panel) the $^8B$ production distribution spectrum,
so that a comparison between the strength of the field and the production
intensity can be directly made.

\begin{figure}[h]
\setlength{\unitlength}{1cm}
\begin{center}
%\hspace*{-1.8cm}
\hspace*{-1.6cm}
\epsfig{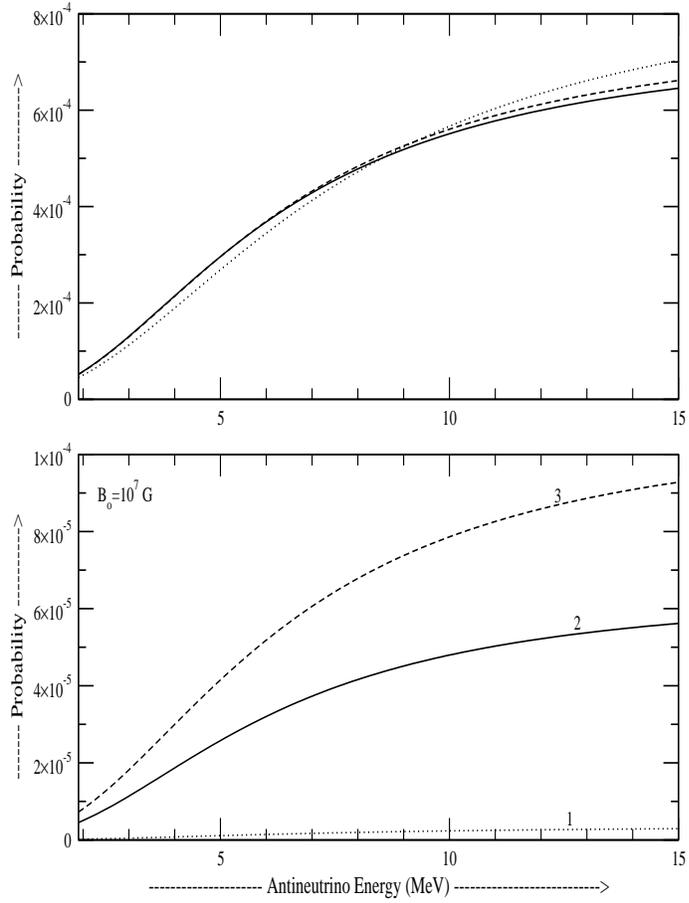}
\end{center}
\caption{ \it Antineutrino probabilities for solar field profiles 1, 2
and 3. Upper panel: the peak field is chosen in each case so as to produce 
the same event rate in KamLAND, (see the main text). Lower panel: the same
value of the peak field ($B_0=10^7G$) is seen in each case to lead to 
probabilities of quite different magnitudes.} 
\label{fig2}
\end{figure}

The antineutrino production probabilities as a function of energy for
each of these profiles are given in fig. \ref{fig2}. In the first panel,
the values of the peak field are chosen so as to produce a fixed number
of events. In this case the
probability curves differ only slightly in their shapes while their 
normalizations 
are the same. The curves are in any case similar to 
the SFP survival probability ones \cite{Pulido:1999xp} in 
the same energy range. 
In the second panel of fig. \ref{fig2} the antineutrino 
probabilities for a
common value of the peak field and these three different profiles are shown. 
It is hence apparent from these two graphs how the distribution of the 
magnetic field intensity is determinant for the magnitude of the 
antineutrino probability, but not for its shape. One important reason for 
this behavior is that we have integrated the antineutrino probability 
over the Boron production region.

%%%% ET MODIFICATION
%{\bf 4.} Following the above considerations on the scaling of the solar
%antineutrino flux, $\phi_{\bar\nu}(E)=\phi^0_{\bar\nu}\times f(E)$, one
%can write the KamLAND event rate for solar antineutrinos (eq.(7)) and a
%given profile $B(r)$ as
%\be
%S=Q_0\int_{E_e^0}^{\infty} dE_e \int_{E_m}^{E_M} \epsilon (E^{'}_e) R(E_e,E^{'}_e)
%\sigma(E) \alpha^2 (xB_0)^2 f(E) dE
%\ee
%where $\alpha$ is a dimensionless factor characterizing the field profile and
%satisfying $\phi^0_{\bar\nu}=\alpha^2 (xB_0)^2$ with $B_0$ denoting the respective peak 
%field value and $x^2=1~cm^{-2}s^{-1}G^{-2}$. Introducing a reference profile 
%$B^{0}(r)$, this scaling property can be written as 
%\be
%\frac{S}{S^0}=\left(\frac{\alpha}{\alpha_0}\right)^2\left(\frac{B_{0}}{B_{0}^{0}}
%\right)^2
%\ee     
%with $S^0$ is the event rate for the reference profile with peak field 
%value $B_{0}^0$.

\section{Results and discussion}
%{\bf 4.} 

The antineutrino signal for any magnetic field profile $B(r)$
can be written, 
taking into account the previous formulas and the near invariance of the 
probability shape (see fig. \ref{fig2}), as 
\begin{eqnarray}
S_{\overline{\nu}}[{B(r)}]&=& \alpha S_{\overline{\nu}}^0
\end{eqnarray}
where $S_{\overline{\nu}}^0$ is the antineutrino signal taken at some 
nominal reference value $B_0^0$ for the field at the solar core for a certain 
reference profile $B^0$. This profile dependent parameter $\alpha$, being a ratio
of two event rates given by eq.(7) for different profiles, can thus be
simplified to 
\be
\alpha=\frac{\int \left(\frac{B(r_i)\sin^2\theta(r_i)}{g^{'}_{2}(r_i)}\right)^2 f_B (r_i)dr_i}
{\int \left(\frac{B^{0}(r_i)\sin^2\theta(r_i)}{g^{'}_{2}(r_i)}\right)^2
 f_B (r_i)dr_i}
\label{alpha}
\ee
where the integrals extend over the production region.
%The KamLAND expected signal for an arbitrary reference condition, 
%profile 2 with $B_0^{0}= 10^7\ G$, is shown in fig. \ref{fig3}. 
As we mentioned before, for concreteness we have fixed along 
this discussion  the neutrino magnetic moment $\mu_{\nu}=10^{-12} \mu_B$.

We will now obtain bounds on parameter $\alpha$ and the peak field $B_0$
for each profile derived from KamLAND data,
%In order to obtain these bounds, the observed KamLAND rates are analyzed 
applying Gaussian probabilistic considerations to the global rate in the
whole energy range, $E_{\nu}=(2.6-8.125)~MeV$, and Poissonian considerations to the 
%global rate seen by the experiment and to the individual 
event content in the highest energy bins ($E_e> 6 $ MeV) where KamLAND 
observes zero events. We denote by $S^{0}_{\bar\nu}$ the event rate with
$B_{0}=10^7G$ for each given profile ($S^{0}_{\bar\nu}=S_{\bar\nu}(10^7G)$). 
%We can make a first estimation of the upper bound on the peak field  
%and the solar antineutrino appearance probability 
Taking the number of observed events and subtracting the 
number of events expected from the best-fit oscillation 
solution [($\Delta m^2,\sin^2 2\theta)_{LMA}=(6.9\times 10^{-5}\ eV^2,1)$]
and interpreting this difference as a hypothetical 
signal coming from  solar antineutrinos, we have
\begin{eqnarray}
S^{sun}_{\overline{\nu}}&=& 
S_{obs}-S_{react}(LMA).
\end{eqnarray}
Inserting \cite{klstony} $S_{obs}=54.3\pm 7.5 $ and
$S_{react}(LMA)= 49\pm 1.3 $, we obtain
$ S_{obs}-S_{react}=\alpha
S^{0}_{\bar\nu} < 17.8\  (20.2)$ at 90 (95)\% CL. 
Within each specific profile it is seen from (\ref{alpha}) that the quantity 
$\alpha$ is simplified to $\alpha=(B_0/10^7G)^2$, so that the previous
inequality becomes  
\be
B_{0}^2<\frac{S^{sun}_{\overline{\nu}}}{S^{0}_{\bar\nu}}(10^7G)^2.
\ee
In this way we can derive for each given profile an upper bound on $B_0$.
The quantity $S^{0}_{\bar\nu}$ for profiles 1, 2 and 3 
and the respective upper bounds on $B_0$ are shown in table 1.  
These upper limits can be cast in a more general way if do not fix
the neutrino magnetic moment. To this end we will consider an arbitrary
reference value $\mu^{0}_{\nu}=10^{-12}\mu_{B}$. Then within each profile, $\alpha=
(\mu_{\nu}B_0/\mu^{0}_{\nu}~10^7G)^2$, where in the numerator and denominator
we have respectively the peak field value and some reference peak field 
value of the same profile. In the same manner as before we can derive the upper
bounds on $\mu_{\nu} B_0$ which are also shown in table 1.  

\vspace{0.6cm}

From the definition of $\alpha$ (\ref{alpha})  
it follows that the upper bounds on the antineutrino flux are independent of the 
field profile. These turn out to be $\phi_{\bar\nu}<0.0034\phi(^8B)$ and 
$\phi_{\bar\nu}<0.0038\phi(^8B)$ for 90 and 95\% CL respectively.

We can similarly and independently apply Poisson statistics to the five highest 
energy bins of the KamLAND experiment. No events are observed in this region and the
expected signal from oscillating neutrinos with LMA parameters is negligibly
small. We use the fact that the sum of Poisson variables of mean $\mu_i$ is
itself a Poisson variable of mean $\sum \mu_i$. The background (here the reactor
antineutrinos) and the signal (the solar antineutrinos) are assumed to be 
independent Poisson random variables with known means.
If no events are observed and in particular no background is observed, the
unified intervals \cite{cousins,Hagiwara:fs} $[0,\epsilon_{CL}]$ are $[0,2.44]$
at 90\% CL and $[0,3.09]$ at 95\% CL.

From here, we obtain $\alpha S_{\bar\nu}^{0} < \epsilon_{CL}$ or 
$\alpha < \epsilon_{CL}/S_{\bar\nu}^{0}$. Hence, as in the previous case,
we have
\be
B_{0}^2<\frac{\epsilon_{CL}}{S^{0}_{\bar\nu}}(10^7G)^2.
\ee
Using the expected number of events in the first 145 days of data taking 
and in this energy range $(6-8.125)~MeV$, we have derived upper bounds on 
$B_0$ (90 and 95\% CL) for all three profiles. They are shown in
table 2 along with the upper bounds on $\mu_{\nu} B_0$ taking $\mu_{\nu}$
as a free parameter.  
The antineutrino flux upper bounds are now $\phi_{\bar\nu}<0.0049\phi(^8B)$
$\phi_{\bar\nu}<0.0055\phi(^8B)$ at 90 and 95\% CL respectively. The KamLAND
expected signal for an arbitrary field profile corresponding to 95\% CL is
shown in fig. 3. 

The differences in magnitude among the bounds on $B_0$ and $\mu_{\nu}B_0$ 
presented in tables 1 and 2 for the different profiles 
are easy to understand. In fact, recalling that the $^8B$ production zone 
peaks at 5\% of the solar radius and becomes negligible at 
approximately 15\% (fig. \ref{fig1}), then
in order to generate a sizeable antineutrino flux, the magnetic 
field intensity should lie relatively close to its maximum in 
the range where the neutrino production is peaked. Thus for 
profile 1 the value of $B_0$ required to 
produce the same signal is considerably larger than for the other two,
while profile 3 is the most efficient one for antineutrino production.  

%These upper limits can be cast in a slightly more general 
%way. If we do not fix the value of the neutrino magnetic moment, then 
%$(\alpha=\mu B_0/ \mu_\nu^0 B_0^0)^2$ and limits above can be translated 
%into an upper limit   $\mu B_0 < 7.02 \times 10^{-19}$ MeV at 95\% CL.
%If we consider a fixed value for the value of the magnetic field at 
%the sun center, for example $B_0^0=10^7\ G$, we trivially obtain 
%an upper limit on the neutrino transition magnetic moment  
%$\mu_\nu < 5.95\times 10^{-12}\ \mu_B $ at 95\% CL, a value of the 
%same order of the astrophysical limits
%\cite{Raffelt:pj}.

As referred to above, for different field profiles the probability curves will
differ only slightly in their shape if they lead to the same number of events.
In other words, for a given number of events the probability curves are essentially
the same, regardless of the field profile, a fact illustrated in fig. \ref{fig2}. As a
consequence, the energy spectrum of the expected solar antineutrino flux will
be nearly the same for any profile. In fig. \ref{fig4} we plot this profile independent 
spectrum together with the $^8 B$ one \cite{Bahc}, so that a comparison can be 
made showing the shift in the peak and the distortion introduced. 

\section{Conclusions}

To conclude, now that SFP is ruled out as a dominant effect for the solar
neutrino deficit, it is important to investigate its still remaining possible
signature in the solar neutrino signal, namely an observable $\bar\nu_e$ flux.
Our main conclusion is that, from the antineutrino production model expound
here, an upper bound on the solar antineutrino flux can be derived, namely 
$\phi_{\bar\nu}<3.8\times 10^{-3}\phi(^8B)$ and $\phi_{\bar\nu}<5.5\times 
10^{-3}\phi(^8B)$ at 95\% CL, assuming respectively Gaussian or Poissonian 
statistics. For 90\% CL we found $\phi_{\bar\nu}<3.4\times 10^{-3}\phi(^8B)$
and $\phi_{\bar\nu}<4.9\times 10^{-3}\phi(^8B)$ which shows an improvement 
relative to previously existing bounds from LSD \cite{Aglietta} by a
factor of 3-5. These are independent of the detailed magnetic field profile 
in the core and radiative zone and the energy spectrum of this flux is also 
found to be profile independent. 
We also derive upper bounds on the peak field value which are 
uniquely determined for a fixed solar field profile. In the most
efficient antineutrino producing case (profile 3), we get  (95\% CL)
an upper limit on the product of the neutrino magnetic moment by the  
solar field $\mu_{\nu} B \leq 2.8\times 10^{-19}$ MeV or  
$B_0 \leq 4.9 \times 10^7 G$  for $\mu_\nu=10^{-12}\mu_B$.
A recent study of the magnetic field in the radiative zone of the sun 
has provided upper bounds of (3-7) MG \cite{Friedland:2002is} in that region in
the vicinity of $0.2~R_S$ which are independent
of any neutrino magnetic moment. Therefore we can use them in 
conjunction with our results to obtain a limit on $\mu_{\nu}$. 
Using $B_0\sim 3-7 MG$, we get from the results for profiles 1-3:  
$\mu~\lsim~0.7-9.6 \times 10^{-12}\mu_B$.
Moreover, from the limits obtained in this work, if the 'true' solar profile resembles 
either a profile like 1 or 3, this criterion implies that SFP cannot be experimentally 
traced in the next few years, since the peak field value must be substantially 
reduced in order to comply with this upper bound, thus leading to a much too 
small antineutrino probability to provide an observable event rate \footnote{Recall 
that the antineutrino probability is proportional to $(\mu_{\nu}B)^2$.}.
On the other hand, for a profile like 2 or in general any one resembling a dipole
field, SFP could possibly be visible.  

\vspace{0.5cm}  
\noindent

{\em Acknowledgements. The work of BCC  was supported by Funda\c{c}\~{a}o para a Ci\^{e}ncia e a 
Tecnologia through the grant SFRH/BPD/5719/2001.
E.T-L  acknowledges many useful conversations with 
P. Aliani, M. Picariello and V. Antonelli, the hospitality of the 
CFIF (Lisboa)
and the  financial  support of the  Spanish CYCIT  funding agency. 
}

\newpage

\begin{table}
\begin{tabular}{||c|c|c|c|c|c||} \hline \hline
     &     &      &     &     &    \\[-0.4cm]
 Profile & $S^{0}_{\bar\nu}(10^7G)$ & $B_{0}(90\% CL)$ & $B_{0}(95\% CL)$ & $\mu_{\nu}B_0(90\% CL)$ & $\mu_{\nu}B_0(95\% CL)$  \\
 & & $G$ & $G$ & $MeV$ & $MeV$  \\[0.1cm]  \hline
1.& $0.006$ & $5.27\times 10^8$ & $5.62\times 10^8$ & $3.05\times 10^{-18}$ & $3.25\times 10^{-18}$\\
2.& $0.137$ & $1.14\times 10^8$ & $1.21\times 10^8$ & $6.60\times 10^{-19}$ & $7.04\times 10^{-19}$\\
3.& $0.224$ & $8.92\times 10^7$ & $9.50\times 10^7$ & $5.16\times 10^{-19}$ & $5.50\times 10^{-19}$\\[0.1cm]
  \hline
\end{tabular}  
\caption{\it  Solar antineutrino event rates, upper bounds on the peak field value
for $\mu_{\nu}=10^{-12}\mu_B$ and on $\mu_{\nu} B_0$ for arbitrary $\mu_{\nu}$ and $ B_0$,
assuming Gaussian statistics in the whole KamLAND spectrum.}
\end{table}

\begin{table}
\begin{tabular}{||c|c|c|c|c|c||} \hline \hline
     &     &      &     &     &    \\[-0.4cm]
 Profile & $S^{0}_{\bar\nu}(10^7G)$ & $B_{0}(90\% CL)$ & $B_{0}(95\% CL)$ & $\mu_{\nu}B_0(90\% CL)$ & $\mu_{\nu}B_0(95\% CL)$  \\
 & & $G$ & $G$ & $MeV$ & $MeV$  \\[0.1cm]  \hline
1.& $0.004$ & $2.53\times 10^8$ & $2.85\times 10^8$ & $1.47\times 10^{-18}$ & $1.65\times 10^{-18}$\\
2.& $0.079$ & $5.56\times 10^7$ & $6.25\times 10^7$ & $3.22\times 10^{-19}$ & $3.62\times 10^{-19}$\\
3.& $0.130$ & $4.34\times 10^7$ & $4.88\times 10^7$ & $2.51\times 10^{-19}$ & $2.82\times 10^{-19}$\\[0.1cm]
   \hline
\end{tabular}                
\caption{\it  Same as table 1 assuming Poissonian statitics in the KamLAND
energy range $E_e=(6-8.125)~MeV$.}
\end{table}

\clearpage

%%\begin{samepage}

\begin{figure}[h]
\setlength{\unitlength}{1cm}
\begin{center}
%\hspace*{-1.8cm}
\hspace*{-1.6cm}
\epsfig{file=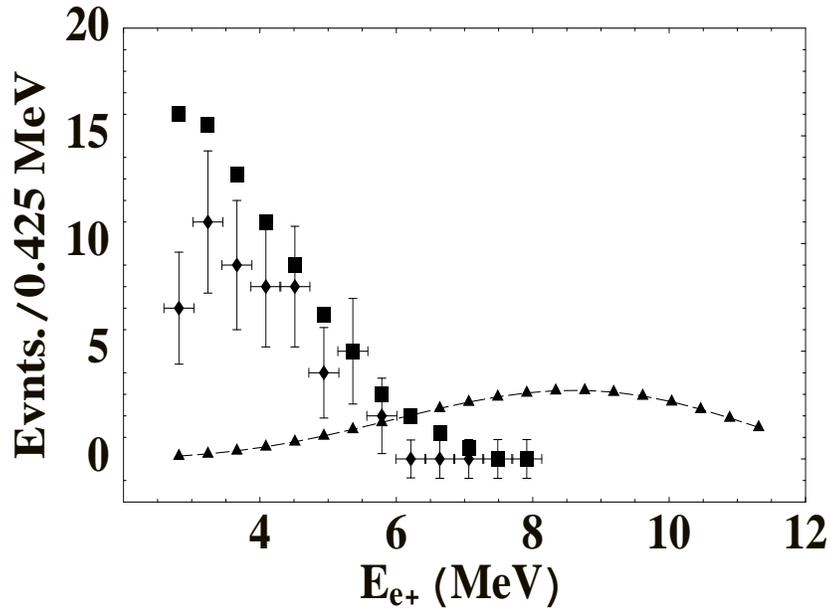,height=9.0cm,width=12.0cm,angle=0}
\end{center}
\caption{ \it 
The solid squares represent the MC expectation of 
the KamLAND positron spectrum from reactor antineutrinos with no 
oscillations and the points with error bars represent the measured
spectrum (from Fig.5 in Ref.\protect\cite{Eguchi:2002dm}).
Solid triangles represent the positron spectrum from solar
antineutrinos (multiplied by 5) assuming profile 3 with
peak field given by its 95\% CL upper limit ($B_0=4.88\times10^7G$).
All curves refer to the same time exposure of 145 days.} 
%The KamLAND  positron spectra from reactor antineutrinos 
%(from Fig.2 in Ref.\protect\cite{Eguchi:2002dm}):
%measured (145.5 days), MC expectations in absence 
%of oscillations respectively points with error-bars, triangles.
%The ``solar'' positron spectrum  (black solid squares) 
%obtained assuming the profile 1 (multiplied by 10$^2$).}
\label{fig3}
\end{figure}

\begin{figure}[h]
\setlength{\unitlength}{1cm}
\begin{center}
%\hspace*{-1.8cm}
\hspace*{-1.6cm}
\epsfig{file=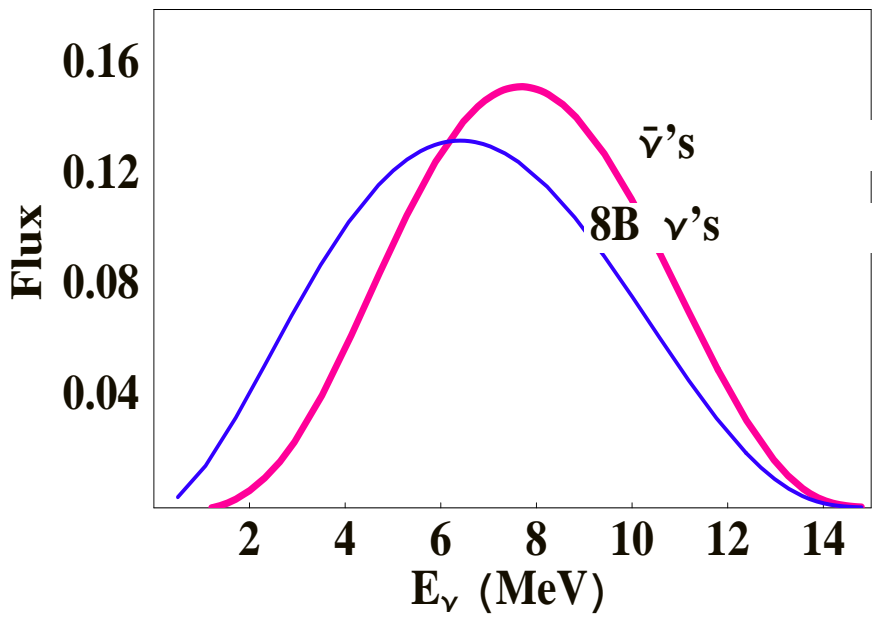,height=9.0cm,width=12.0cm,angle=0}
\end{center}
\caption{ \it The expected solar antineutrino spectrum and the 
$^8 B$ neutrino one \cite{Bahc}, both normalized to unity, showing the 
peak shift and the distortion introduced by the antineutrino probability.}
\label{fig4}
\end{figure}

\end{document}